\documentclass[useAMS,usenatbib]{mn2e}
\usepackage{graphicx,natbib,color,multirow,amsmath}
\definecolor{titlecol}{rgb}{0,0,1}
\definecolor{titlecol2}{rgb}{0,0.65,0}
\definecolor{titlecol3}{rgb}{0.99,0.4,0.} 

\def\changed    {}
\def\changedtwo    {}

\def\mmsun	{\rm{M}_{\odot}}

\def\diskparent {\textsc{disk-parent}}
\def\barparent {\textsc{bar-parent}}
\def\disklz {\textsc{disk-lz}}
\def\barlz {\textsc{bar-lz}}
\def\smoothlz {\textsc{smooth-lz}}
\def\pbar {p_{s,bar}}

\def\lesssim{\mathrel{\hbox{\rlap{\hbox{\lower3pt\hbox{$\sim$}}}\hbox{\raise2pt\hbox{$<$}}}}}
\def\gtrsim{\mathrel{\hbox{\rlap{\hbox{\lower3pt\hbox{$\sim$}}}\hbox{\raise2pt\hbox{$>$}}}}}

\begin{document}

\title[Bars and Bar Fractions in Galaxy Zoo: CANDELS]{Galaxy Zoo: CANDELS Barred Disks and Bar Fractions\thanks{This publication has been made possible by the participation of more than 200,000 volunteers in the Galaxy Zoo project. Their contributions are
individually acknowledged at http://authors.galaxyzoo.org/ .} } 

\author[Simmons et al.]{\parbox[t]{16cm}{B. D. Simmons$^{1}\thanks{E-mail: brooke.simmons@astro.ox.ac.uk}$, 
Thomas Melvin$^{2}$, 
Chris Lintott$^{1,3}$, 
Karen L. Masters$^{2,4}$, 
Kyle W. Willett$^{5}$, 
William C. Keel$^{6}$, 
R. J. Smethurst$^{1}$, 
Edmond Cheung$^{7}$, 
Robert C. Nichol$^{2,4}$, 
Kevin Schawinski$^{8}$, 
Michael Rutkowski$^{5}$,
Jeyhan S. Kartaltepe$^{9,}$\footnote{Hubble Fellow}, 
Eric F. Bell$^{10}$,
Kevin R. V. Casteels$^{11}$, 
Christopher J. Conselice$^{12}$,
Omar Almaini$^{12}$, 
Henry C. Ferguson$^{13}$,
Lucy Fortson$^{5}$, 
William Hartley$^{12,8}$, 
Dale Kocevski$^{14}$,
Anton M. Koekemoer$^{13}$,
Daniel H. McIntosh$^{15}$,
Alice Mortlock$^{12}$, 
Jeffrey A. Newman$^{16}$,
Jamie Ownsworth$^{12}$, 
Steven Bamford$^{12}$,
Tomas Dahlen$^{13}$,
Sandra M. Faber$^{17}$,
Steven L. Finkelstein$^{18}$, 
Adriano Fontana$^{19}$,
Audrey Galametz$^{19}$,
N. A. Grogin$^{13}$,
Ruth Gr\"utzbauch$^{12, 20}$,
Yicheng Guo$^{17}$,
Boris H\"au\ss ler$^{12,21,1}$,
Kian J. Jek$^{22}$,
Sugata Kaviraj$^{21}$,
Ray A. Lucas$^{13}$,
Michael Peth$^{23}$,
Mara Salvato$^{24}$,
Tommy Wiklind$^{25}$,
Stijn Wuyts$^{24}$
\vspace{0.1in} }\\
$^{1}$Oxford Astrophysics, Denys Wilkinson Building, Keble Road, Oxford OX1 3RH, UK\\
$^{2}$Institute of Cosmology \& Gravitation, University of Portsmouth, Dennis Sciama Building, Portsmouth PO1 3FX, UK\\
$^{3}$Adler Planetarium, 1300 S. Lake Shore Drive, Chicago, IL 60605, USA\\
$^{4}$SEPnet,\thanks{www.sepnet.ac.uk} South East Physics Network\\
$^{5}$School of Physics and Astronomy, University of Minnesota, 116 Church St. SE, Minneapolis, MN 55455, USA\\
$^{6}$Department of Physics and Astronomy, University of Alabama, Box 870324, Tuscaloosa, AL 35487, USA\\
$^{7}$Department of Astronomy and Astrophysics, 1156 High Street, University of California, Santa Cruz, CA 95064, USA\\
$^{8}$Institute for Astronomy, ETH Z\"urich, Wolfgang-Pauli-Strasse 27, CH-8093 Z\"urich, Switzerland\\
$^{9}$National Optical Astronomy Observatory, 950 N. Cherry Ave., Tucson, AZ, 85719, USA\\
$^{10}$Department of Astronomy, University of Michigan, Ann Arbor, MI 48104, USA\\
$^{11}$Institut de Cincies del Cosmos. Universitat de Barcelona (UB-IEEC), Mart i Franqus 1, E-08028 Barcelona, Spain\\
$^{12}$School of Physics \& Astronomy, University of Nottingham, Nottingham NG7 2RD\\
$^{13}$Space Telescope Science Institute, 3700 San Martin Drive, Baltimore, MD 21218\\
$^{14}$Department of Physics and Astronomy, University of Kentucky, Lexington, KY 40506, USA\\
$^{15}$Department of Physics {\changed and Astronomy}, University of Missouri-Kansas City, 5110 Rockhill Road, Kansas City, MO 64110, USA\\
$^{16}$Department of Physics and Astronomy \& PITT PACC, University of Pittsburgh, Pittsburgh, PA 15217, USA\\
$^{17}$UC Observatories/Lick Observatory and Department of Astronomy and Astrophysics, University of California, Santa Cruz, CA 95064, USA\\
$^{18}$Department of Astronomy, The University of Texas at Austin, Austin, TX 78712, USA\\
$^{19}$INAF-Osservatorio Astronomico di Roma, Via Frascati 33, I-00040, Monteporzio, Italy\\
$^{20}$Centre for Astronomy and Astrophysics, University of Lisbon, P-1349-018 Lisbon, Portugal\\
$^{21}$Centre for Astrophysics Research, University of Hertfordshire, College Lane, Hatfield AL10 9AB, UK\\
$^{22}$Galaxy Zoo Volunteer\\
$^{23}$Department of Physics and Astronomy, The Johns Hopkins University, Baltimore, MD 21218, USA\\
$^{24}$Max-Planck-Institut f{\"u}r extraterrestrische Physik, Giessenbachstrasse 1, DÐ85748 Garching bei M{\"u}nchen, Germany\\
$^{25}$European Southern Observatory/Joint ALMA Observatory, 3107 Alonso de Cordova, Santiago, Chile\\
   }

\maketitle
  
\label{firstpage}
  
\clearpage

\begin{abstract}

The formation of bars in disk galaxies is a tracer of the dynamical maturity of the population. Previous studies have found that the incidence of bars in disks decreases from the local Universe to $z \sim 1${\changed , and by $z > 1$} simulations predict that bar features in dynamically mature disks should be extremely rare. Here we report the discovery of strong barred structures in massive disk galaxies at $z \sim 1.5$ in deep rest-frame optical images from CANDELS. From within a sample of 876 disk galaxies identified by visual classification in Galaxy Zoo, we identify 123 barred galaxies. Selecting a sub-sample within the same region of the evolving galaxy luminosity function {\changed  (brighter than $ L^*$)}, we find that the bar fraction across the redshift range $0.5 \leq z \leq 2$ ({\changed $f_{bar} = 10.7^{+6.3}_{-3.5}\%$} after correcting for incompleteness) does not significantly evolve. We discuss the implications of this discovery in the context of existing simulations and our current understanding of the way disk galaxies have evolved over the last 11 billion years.

  \end{abstract}
  

%
%
\section{Introduction}
%
%

Large-scale galactic stellar bars {\changed are thought to form} within dynamically cold, rotationally supported disks \citep{athanassoula05b, combes09,athanassoula13,sellwood13}. Thus{\changed ,} the evolution of the fraction of disk galaxies with bar features traces the overall evolution of disk galaxy dynamics. Locally, bars are present in $\sim 25 - 50\%$ of disk galaxies \citep[{\changed where bars are classified either visually, from Fourier analysis, or from examining elliptical isophotes;} e.g.][]{odewahn96a, d_elmegreen04b, aguerri09, nair10, masters11a, cheung13}, with their abundance steadily decreasing to $\sim$10\% of disk galaxies at $z \sim 1$ \citep{abraham99, d_elmegreen04b, d_elmegreen05, sheth08, melvin14}. 

The lower incidence of bars at higher redshifts may be in part be due to the increased incidence of mergers and galaxy interactions \citep{conselice03b, lotz11, casteels13}, which disrupt and heat disks, destroying or preventing the formation of bars. {\changed More generally, disk galaxies at $z \sim 1$ tend to be less dynamically ``settled'' than their more local counterparts, with a lower rotation velocity compared to velocity dispersion as redshift increases \citep{forsterschreiber11a,kassin12}. The redshift dependence of bar fractions} may also be related to the expected increase in disk gas fraction with redshift; this has been {\changed observed directly} via the increased $M_{gas}/M_{star}$ from CO observations (e.g., \citeauthor{tacconi10} \citeyear{tacconi10}, \citeyear{tacconi13}; for a detailed review, see \citeauthor{carilli13} \citeyear{carilli13}). The presence of a bar in a galaxy is anti-correlated with specific star formation rate \citep{cheung13} and disk gas fraction \citep{masters12a}, in agreement with theoretical predictions \citep{friedli93, berentzen07, villavargas10, athanassoula13}, {\changed although a high gas fraction does not entirely preclude the existence of a bar \citep{nair10, masters12a}}. 

The current theoretical understanding of bar fraction evolution suggests that disk galaxies at $z > 1$ {\changed may be} too dynamically hot to form bars{\changed :} \citet*{kraljic12} find no observable bars within a simulated sample of galaxies at $z \sim 1.5$. {\changed However, other} simulations explore the impact of tidal heating and galaxy harassment, which can either inhibit bar formation \emph{or} promote it, depending on mass {\changed \citep{noguchi88, moore96, skibba12}. Testing the viability of the proposed mechanisms responsible for the redshift dependence of bar fractions} requires high-resolution imaging over a large area of the sky to observe statistically significant samples in multiple redshift bins and adequate spatial resolution to resolve galactic-scale bars in the rest-frame optical \citep[since the detectability of bars decreases rapidly blueward of the 4000~\AA\ break;][]{sheth08}.

These observing requirements currently limit studies of disk populations via bar fractions to surveys with the \emph{Hubble Space Telescope (HST)}. Previous studies have used the optical cameras on \emph{HST} to examine bar fractions to $z \sim 1$. In this paper, we present {\changed the} first results from Galaxy Zoo morphological classifications of galaxies imaged by the Cosmic Assembly Near-Infrared Deep Extragalactic Legacy Survey \citep[CANDELS;][]{grogin11,koekemoer11}, which uses \emph{HST's} near-infrared Wide-Field Camera 3 (WFC3) {\changed and which allows us to probe the bar fractions of galaxies with $L > L^*$ out to $z \sim 2$}.

In Section 2 we describe our sample selection, including a summary of Galaxy Zoo classifications of CANDELS galaxies and how disks and bars are selected. We also explore any potential biases that may affect our results. We present our results in Section 3, with a discussion including comparison to simulated predictions in Section 4, and a summary in Section 5. Throughout
this paper we use the AB magnitude system, and where necessary we adopt a cosmology consistent with $\Lambda$CDM, with $H_{\rm 0}=70~{\rm
km~s^{-1}}$Mpc$^{\rm -1}$, $\Omega_{\rm m}=0.3$ and $\Omega_{\rm \Lambda}=0.7$ \citep{bennett13}.

%
%
\section{Data}\label{sec:data}
%
%

\subsection{CANDELS}\label{candelsdata}

The Cosmic Assembly Near-infrared Extragalactic Legacy Survey \citep[CANDELS;][]{grogin11,koekemoer11} is an \emph{HST} Treasury program combining optical and near-infrared imaging from the Advanced Camera for Surveys (ACS) and Wide Field Camera 3 (infrared channel; WFC3/IR) across five well-studied survey fields {\changed (GOODS-North and -South, \citeauthor{giavalisco04} \citeyear{giavalisco04}; EGS, \citeauthor{davis07} \citeyear{davis07}; UDS, \citeauthor{lawrence07} \citeyear{lawrence07}, \citeauthor{cirasuolo07} \citeyear{cirasuolo07}; and COSMOS, \citeauthor{scoville07} \citeyear{scoville07}) using a two-tiered ``deep'' and ``wide'' approach. Each of the wide fields (UDS, COSMOS, EGS and flanking fields to the GOODS-S and GOODS-N deep fields) are imaged over 2 orbits in WFC3/IR, split in a 2:1 ratio between filters F160W and F125W, respectively, with parallel exposures in F606W and F814W using ACS. Each of the deep fields (GOODS-S and GOODS-N) are imaged over at least 4 orbits each in both the F160W and F125W filters and 3 orbits in the F105W filter, with ACS exposures in F606W and F814W in parallel.} These are reduced and combined to produce a single mosaic for each field {\changed in each band}, with drizzled resolutions of $0.03^{\prime\prime}$ and $0.06^{\prime\prime}$ per pixel for ACS and WFC3/IR, respectively \citep[a process described in detail by][]{koekemoer11}.

Here we use the CANDELS ACS and WFC3/IR images from within the COSMOS, GOODS-South, and UDS fields for which raw classifications from the Galaxy Zoo project are presently available. The WFC3/IR observations of these fields cover approximately {\changed 0.15} square degrees combined. The Galaxy Zoo classifications are based on colour images created using {\changed the \citet{lupton04} asinh stretch method with resolution-matched} WFC3 F160W, F125W, and ACS F814W as red, green and blue channels {\changed respectively.} Some of the colour images use ACS data that was observed during previous surveys \citep{giavalisco04,scoville07,koekemoer07,davis07} and re-analysed by the CANDELS pipeline.

\subsection{Classifications}\label{sec:gzdata}

Galaxy Zoo provides quantified visual morphologies by obtaining multiple independent classifications for each galaxy. Beginning in 2007, more than 1,000,000 galaxy images in total from the Sloan Digital Sky Survey and the \emph{HST} have each been classified by typically $\sim 40$ independent volunteers via a web interface\footnote{zoo4.galaxyzoo.org}. The initial version of the project \citep{lintott08,lintott11} asked a single question per galaxy (whether the galaxy was spiral or elliptical). Subsequent versions have collected more detailed morphological information, including finer sub-structures of disk galaxies such as bulge strength and bars, via a tiered classification tree \citep[e.g.,][]{willett13,melvin14}. {\changed All previous Galaxy Zoo projects have incorporated extensive analysis of volunteer classifications to measure classification accuracy and bias and compute user weightings \citep[for a detailed description of debiasing and consistency-based user weighting, see Section 3 of][]{willett13}. The classifications are highly accurate and the high number of independent classifications per galaxy has enabled a diverse range of investigations in the overall field of galaxy evolution \citep[e.g.,][]{darg10b,darg10a,masters11a,skibba12,casteels13}.}

This work uses classifications collected during the fourth release of Galaxy Zoo, specifically of 49,555 images from the COSMOS, GOODS-South, and UDS fields in the CANDELS survey (hereafter GZ-CANDELS). The dataset was initially composed of all sources having $F160W$ ($H$) apparent magnitude $< 25.5${\changed . {\changedtwo This limit is considerably brighter than the expected $5 \sigma$ extended-source detection limits reported in \citet[][]{grogin11}.}
 Within this sample,} 58\% of sources have $25.5 < H < 24.5$, and 31\% of sources have $H < 23.5$. {\changed (}We note that this brighter sub-sample includes 95\% of galaxies later selected as ``featured'' galaxies{\changed ;} Section \ref{sec:sample}). 

{\changed Several months after the launch of GZ-CANDELS, an initial analysis motivated by community\footnote{talk.galaxyzoo.org} tags of sources considered too faint to classify resulted in the application of systematic cuts in magnitude-surface-brightness space and the} early retirement of 1,555 point-like sources and 11,837 faint, low-surface brightness galaxies without resolvable fine features. Although the project is still ongoing, as of the date of this analysis each of the remaining objects has received at least 40 independent classifications {\changed (mean number 43; maximum 81). For each source classified by volunteers in GZ-CANDELS, all independent classifications are combined to produce ``vote percentages'', where a vote percentage $p$ for a given answer to a given question in the classification tree is the number of votes for that answer divided by the number of classifiers who answered the question.}

\begin{figure*}
\includegraphics[scale=0.273]{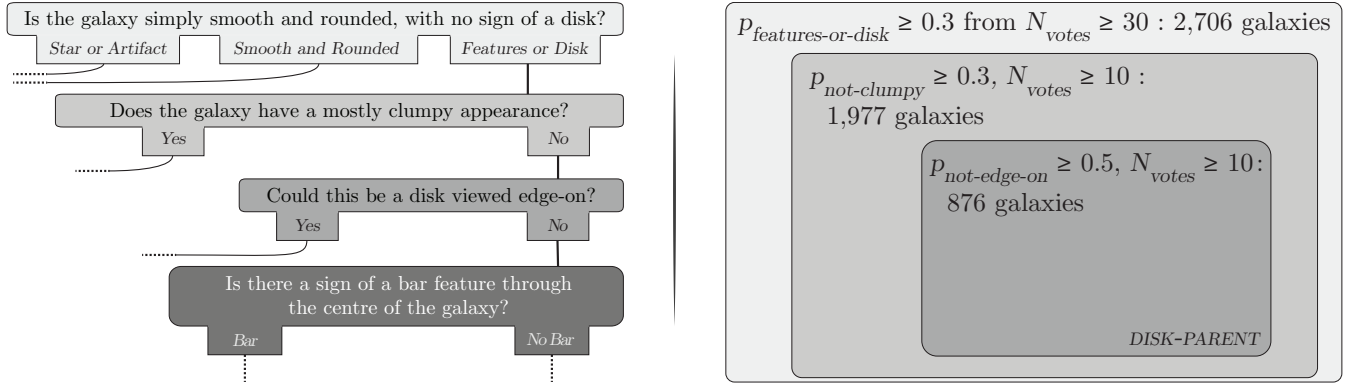}
\caption{
\emph{Left:} Partial Galaxy Zoo: CANDELS classification tree, starting with the first question (top) and leading to the bar feature question. There are 17 questions total in the tree; the bar question is a 4th-tier task. \emph{Right:} Selection of the featured, not-edge-on disk galaxy sample (876 galaxies{\changed , hereafter called the \diskparent\ sample}) in GZ-CANDELS; relative box areas are scaled to the sample sizes. This selection was made independently of restrictions on redshift or luminosity (a full description of the sample selection is given in Section \ref{sec:sample}). Eight independent classifiers subsequently examined each of the 876 {\changed \diskparent } galaxies for evidence of a bar. 
}
\label{fig:sampleselection}
\end{figure*}

The classification tree used for GZ-CANDELS (B. Simmons et al., in preparation; see Figure \ref{fig:sampleselection} for the portion relevant here) first asks volunteers to choose whether a galaxy is mostly smooth, has features, or is a star/artifact. The bar classification question (``Is there a sign of a bar feature through the centre of the galaxy?'') is reached once a volunteer has chosen ``Features or Disk'' as an answer to the first question and has subsequently said the galaxy does \emph{not} have a mostly clumpy appearance, nor is it an edge-on disk. The bar classification is therefore a fourth-tier task, and the number of volunteers per galaxy who answer the bar question varies depending on responses to the earlier tasks. {\changed We discuss the details of the disk and barred disk galaxy sample selections based on the tiered questions in the tree in Section \ref{sec:sample}.}

{\changedtwo We note that an independent effort is underway to collect at least 3 visual classifications per galaxy from CANDELS team members \citep[classifications for one field have been published;][]{kartaltepe14}. These classifications use a different approach to the decision tree method used by Galaxy Zoo; we defer a full analysis between these classifications and Galaxy Zoo results to our upcoming data release paper.}

\begin{figure*}
\includegraphics[scale=1.0]{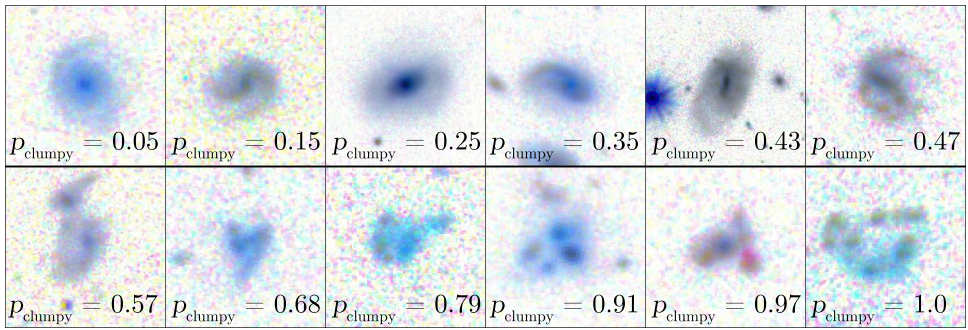}
\caption{
{\changed Examples of galaxies in GZ-CANDELS with different vote percentages for the question ``Does the galaxy have a mostly clumpy appearance?'' Each galaxy is labelled with its clumpy vote percentage, where $p_{clumpy}$ indicates the fraction of classifiers who answered ``Yes'' to the question. (For comparison to the selection described in Section \ref{sec:sample} and Figure \ref{fig:sampleselection}, note that $p_{not-clumpy} = 1 - p_{clumpy}$.) 
In order to favour inclusiveness of clumpy disks while ensuring enough votes for the subsequent questions along the not-clumpy branch of the classification tree, all galaxies with $p_{clumpy} < 0.7$ are included in the disk sample if they also meet the other selection criteria described in Section \ref{sec:sample}.
}
}
\label{fig:clumpy}
\end{figure*}

\subsection{Redshifts}\label{sec:z}

Each of the fields covered by CANDELS data has considerable ancillary data from previous and ongoing work. In addition to newly-calculated photometric redshifts in CANDELS {\changed \citep[based on a Bayesian approach combining results from multiple different analyses;][]{dahlen13}}, we assemble additional photometric and spectroscopic redshifts from the available literature. For galaxies in the COSMOS field, we combine spectroscopic redshifts from the zCOSMOS project \citep{lilly07} with photometric redshifts from COSMOS \citep{ilbert09} and from the NEWFIRM medium-band survey \citep{whitaker11}. In the GOODS-South field, we use the catalog of \citet{cardamone10b}, who added photometric redshifts based on deep broad- and medium-band data from MuSYC \citep{gawiser06} to available spectroscopic redshifts compiled from multiple sources \citep{balestra10,vanzella08,lefevre04,cimatti02}. In the UDS field, we use available spectroscopic {\changed \citep{simpson12}} and photometric redshifts {\changed \citep{hartley13}}, the latter of which make use of deep multi-wavelength coverage from UKIDSS as well as $J$ and $H$-band magnitudes from CANDELS. Of the 49,555 galaxies originally included in Galaxy Zoo: CANDELS, 46,234 currently have spectroscopic (2,886) or photometric (43,348) redshifts. Where available, agreement between spectroscopic and photometric redshift is generally very good, with {\changed $\Delta z \equiv \sigma_z/(1+z_{spec}) = 0.02$ and $\sim 8\%$ of sources having $\Delta z > 0.2$. The use of photometric redshifts introduces an uncertainty of less than $1\%$ into the population bar fractions discussed in Section \ref{sec:results}. These uncertainties are discussed along with other possible sources of error in Section \ref{sec:uncertainties}.}

\subsection{Sample Selection}\label{sec:sample}

\begin{figure*}
\includegraphics[scale=1.0]{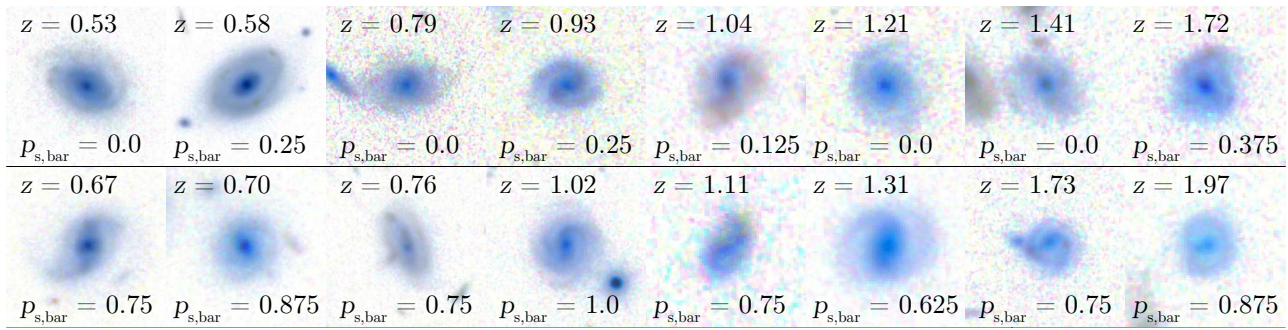}
\caption{
Examples of disk galaxies in GZ-CANDELS whose bar vote percentage $(\pbar)$ places them in the unbarred (top row) and barred (bottom row) sub-samples.
}
\label{fig:gals}
\end{figure*}

A full reduction of the GZ-CANDELS classifications, resulting in a catalog of morphological {\changed measurements that incorporates weighted user votes and adjustments for biases in classifications due to cosmological dimming and resolution effects} for each galaxy, is ongoing. Here we use the raw vote percentages, which have been neither weighted nor debiased. The effects of using raw versus the reduced classifications are twofold. First, the unweighted votes are likely biased in the first question toward an excess of votes for ``Star or Artifact'' (see \citeauthor{willett13} \citeyear{willett13} for a discussion of how inconsistent votes are downweighted in Galaxy Zoo 2, which has a similar {\changed classification} tree). Second, the effects of surface brightness dimming and loss of spatial resolution are not accounted for in the vote percentages, which is potentially a significant effect in a sample extending to $z \sim 2$ in the rest-frame optical. {\changed We minimize the effect of surface brightness dimming via luminosity cuts described in Section \ref{sec:zlumcuts}, and address the lack of user weighting via threshold cuts in classification vote fractions, as described below.}

To favour completeness in the final disk galaxy sample and to minimize the impact of the lack of user weighting, we employ a lower vote percentage threshold when selecting ``featured'' galaxies than is typical when using weighted data. We select as ``featured'' galaxies those where at least 30\% of votes (out of at least 30 volunteers total) were registered for ``Features or Disk''. This selects 2,706 featured galaxies. {\changed For comparison, a more typical threshold for weighted classifications is $p_{features-or-disk} = 0.5$ \citep[e.g.,][]{melvin14}.} After the first question, the user weighting used by previous Galaxy Zoo data reductions {\changed \citep{lintott08, bamford09, willett13}} affects vote percentages by typically no more than a few percent{\changed . We} therefore expect the lack of weighting to have little to no systematic effect on additional vote percentages. 

Subsequent to the featured galaxy selection, we select a sub-sample where at least 30\% of volunteers (where a minimum of 10 answered the question) registered a vote for ``no'' to the question ``Does the galaxy have a mostly clumpy appearance?'' {\changed Figure \ref{fig:clumpy} shows examples of galaxies along the full range of clumpy/not-clumpy vote percentages.  We include the clumpy selection in order to explicitly consider each branch of the classification tree that leads to the bar-feature question, but the threshold is deliberately lower than previous studies \citep{melvin14} to favour inclusiveness of clump-dominated disks that may be more prevalent at higher redshift \citep[e.g.,][]{d_elmegreen04,d_elmegreen05,d_elmegreen07,forsterschreiber11a} while removing galaxies with no apparent underlying disks or whose morphologies preclude evaluation of potential bar features. This selection removes 729 sources in total, leaving 1,977 galaxies. 
However, were we} to ignore the clump-threshold criterion completely, this would only {\changed affect the final sample of disk galaxies} at the 1\% level, due to the subsequent inclination and luminosity selection criteria. Our qualitative results are thus not sensitive to the specific choice of clumpy threshold between $0.1 \leq p_{\mathrm not-clumpy} \leq 0.6$. 

\begin{figure}
\includegraphics[scale=0.41]{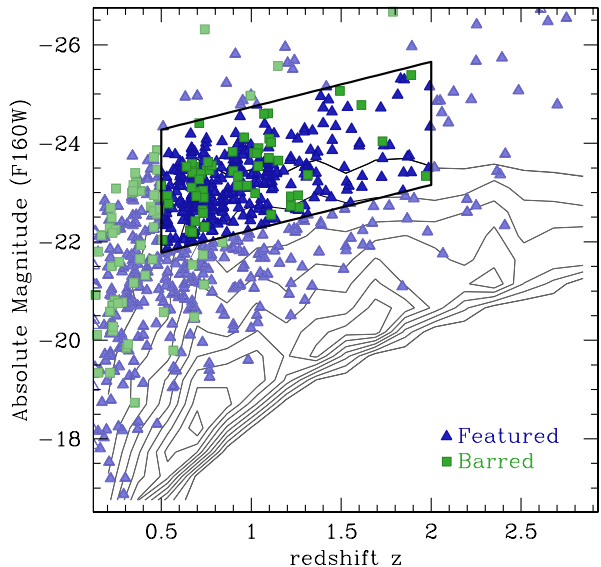}
\caption{
Absolute $H$-band magnitude versus redshift for all sources with $H < 25.5$ (contours in steps of 10\%) and {\changed 876 disks} ({\changed \diskparent\ sample}; blue triangles), of which 123 galaxies show clear evidence of a bar ({\changed \barparent \ sample}; green squares). To facilitate comparison between lookback times, avoid biases due to surface-brightness dimming when calculating bar fractions, and ensure all observed $H$-band flux is redward of the 4000~\AA\ break, we select sub-samples within the same region of the evolving galaxy luminosity function \citep{marchesini12} and $0.5 \leq z \leq 2$ (parallelogram). Within this region there are {\changed 370 disk} galaxies {\changed (\disklz\ sample)}, 56 of which have clear evidence of bars {\changed (\barlz\ sample)}.
}
\label{fig:mag_z}
\end{figure}

{\changed We} also require that 50\% of volunteers (of at least 10) registered a vote for a disk galaxy that is ``not-edge-on''.
This is a deliberately conservative choice to reflect the fact that bars would be invisible in edge on systems (the thresholds used to select disk features are less strict to favour completeness).\footnote{The discussion in Section \ref{sec:results} assumes the bar fraction is the same in edge-on galaxies as face-on {\changed galaxies; an} application of the results to include {\changed strongly} clump-dominated galaxies requires a similar assumption.} This selects a sample of {\changed 876 disk galaxies, within each of} which a bar may be identified, if it exists. {\changed This parent sample of 876 not-edge-on disk galaxies is referred to hereafter as the \diskparent\ sample. As a sanity check on the selection of disks, we examine the \citet{sersic68} indices of the \diskparent\ sample using the parametric fits of \citet{vanderwel12}. We find that the distribution of S\'ersic indices is peaked at $n = 1.4$, with $\sigma_n = 0.6$, fully consistent with a disk-dominated sample \citep[e.g.,][]{haussler07, simmons08}.

Figure} \ref{fig:sampleselection} shows a visual representation of this sample selection, from which a further sub-sample of barred galaxies may be identified. However, approximately 20\% of these 876 galaxies received less than 10 raw votes \emph{total} for the question ``Is there any sign of a bar feature through the centre of the galaxy?'', a consequence of the broad initial selection of featured galaxies and the multiply-branched nature of the classification tree. Because of {\changed this incompleteness and} the lower number of votes per galaxy in the 4$^{\rm{th}}$ tier of the classification tree (the position of the bar question), within the {\changed \diskparent } sample the raw bar fractional vote is statistically useful, but uncertain for individual galaxies.

We therefore elected to supplement the volunteer data with visual classifications from the Galaxy Zoo science team to select the sub-sample of barred disk galaxies. Eight of the authors\footnote{BDS, TM, KWW, WCK, MR, KLM, R{\changed J}S, EC} inspected each of the 876 {\changed \diskparent } galaxies for evidence of a {\changed bar. These} votes were unanimous approximately 60\% of the time, either for a bar feature (23 galaxies) or no bar (512 galaxies). Among galaxies where the science team voted unanimously that a bar is present, the 
{\changed median volunteer bar vote percentage and interquartile range are $0.6_{-0.06}^{+0.17}$.}
Among galaxies where the science team was unanimous that a bar is \emph{not} present, the 
{\changed median volunteer bar vote percentage and interquartile range are $0.1_{-0.1}^{+0.08}$.} 
 The science team and volunteer bar vote percentages {\changed correlate ($r = 0.8${\changedtwo\ with very high significance\footnote{\changedtwo The statistics package {\tt R} reported $p< 2.2 \times 10^{-16}$, which indicates a value smaller than can be precisely reported using floating-point precision.}: 
$p$-value $ \rightarrow  0$})}, 
although the low number of volunteer votes for many objects means the dispersion in the correlation is high. {\changed We therefore choose not to include the incomplete volunteer votes for this question, considering only the science-team classifications in the determination of the bar percentage, $\pbar$.} Following vote percentage thresholds used in previous {\changed studies, we} mark a galaxy as barred if at least half of the science-team classifiers indicated the presence of a bar ($\pbar \geq 0.5$). {\changed This vote fraction threshold has been shown to select strong bars \citep{masters11a,willett13,melvin14}. Using this vote percentage threshold identifies 123 barred disk galaxies (hereafter the \barparent\ sample) from among the \diskparent\ sample of 876 disk galaxies selected from the full GZ-CANDELS sample.}

\subsubsection{{\changed Redshift and Luminosity Selections}}\label{sec:zlumcuts}

The absolute $H$-band magnitudes in the sample are plotted as a function of redshift in Figure \ref{fig:mag_z}. {\changed Within the \diskparent\ sample, 525 galaxies} have redshifts between $0.5 \leq z \leq 2.0${\changed ; {\changed within the \barparent\ sample, 61 galaxies} have $0.5 \leq z \leq 2.0$}. Within this redshift range, all flux {\changed measured in} the WFC3 $H$ band is redward of the 4000~\AA\ break. Examples of barred and unbarred galaxies are shown in Figure \ref{fig:gals}.

To minimize any bias caused by surface-brightness dimming at higher redshifts, we additionally employ a conservative luminosity cut when examining bar fractions, choosing a minimum $H$ absolute magnitude of $-23.15$ at $z = 2${\changed . This} ensures that {\changed galaxy features} can be detected within the sub-sample at all $z < 2$. We note that this is brighter than the knee of the rest-frame-$V$-band luminosity function at this redshift \citep{marchesini12}. In order to examine similar populations across our entire redshift range, we use a redshift-dependent luminosity cut based on selecting the same region of the evolving luminosity function \citep[corrected to observed $H$ band;][]{blanton07,marchesini12}: this selection is shown as a parallelogram shape in Figure \ref{fig:mag_z}. This final cut produces 370 not-edge-on disk galaxies {\changed within these luminosity and redshift bounds}, of which 56 have strong bar signatures. {\changed Hereafter we refer to these as the \disklz\ and \barlz\ samples, respectively.} We note that our results are robust to small variations in the redshift and luminosity thresholds chosen for the sample. For example, our qualitative result does not change if we use a fixed luminosity/stellar mass range.

\subsubsection{Completeness corrections}\label{sec:completeness}

{\changed For galaxies within the luminosity ranges considered here and observed at the depth of the CANDELS images (even the shallower ``wide'' fields), the composition of the final \disklz\ sample is unlikely to be affected by surface brightness dimming. Furthermore, the analysis in this paper is concerned with large-scale, strong galactic bars, which are less affected by surface brightness dimming or the effects of diminishing resolution than weaker features. The result is a conservative selection with respect to feature detection, in the sense that both strong bars in particular and disks with features in general are unlikely to be missed.

However, it is necessary to account for the possibility that a substantial number of rotationally-supported disks with deceptively smooth distributions of light (i.e., disk galaxies that are entirely lacking in `features') might be present in the sample. The presence of such a population would result in our measured bar fractions (Section \ref{sec:results}) being overestimates.

To estimate the maximum contamination from such a population, we examine `smooth' galaxies. In particular, we examine a subsample of all galaxies within our luminosity and redshift cuts (Section \ref{sec:zlumcuts}) with fewer than 30\% of votes in the first question (from a total of at least 30) for either `Features or Disk' or `Star or Artifact' (hereafter the \smoothlz\ sample). We assume this sample has a mix of rotation-dominated and dispersion-dominated galaxies, and we assess the maximum fraction of this sample that could reasonably be a disk population using measurements of axis ratios for these systems (\citeauthor{galametz13} \citeyear{galametz13}; \citeauthor{guo13} \citeyear{guo13}; Fontana et al. 2014, in press; M. Peth et al., in preparation). 

Typical low-redshift disk galaxies have minimum (i.e., edge-on) axis ratios varying from $0.08 \leq (b/a)_{min} \leq 0.2$ \citep[depending on bulge strength; e.g.,][]{padilla08}, and this minimum disk thickness likely increases somewhat for disk galaxies at higher redshift \citep[e.g.,][]{forsterschreiber09,law09}. To account conservatively for the possible thickening of disks, we assume that \emph{all} featureless galaxies with axis ratios $b/a \leq 0.4$ (i.e., ellipticities $\epsilon \geq 0.6$) are disk galaxies. Assuming these are part of a randomly-oriented population of disks, we use their expected distribution of axis ratios (\citeauthor*{lambas92} \citeyear{lambas92}; \citeauthor{binneymerrifield} \citeyear{binneymerrifield}; \citeauthor{padilla08} \citeyear{padilla08}; \citeauthor{law12b} \citeyear{law12b}) to constrain the fraction of the \smoothlz\ sample that is composed of this hypothetical disk population. This fraction is $\approx 19\%$ for the full sample, and generally increases with redshift within our limits between 15\% and 25\%. It should be noted this is likely an overestimate of the contamination, as dispersion-dominated early-type galaxies with smooth light profiles and low axis ratios certainly exist \citep[][]{emsellem11}.

In order to account for this possible contamination, we then apply these fractions to that part of the \smoothlz\ sample that is consistent with the not-edge-on selection described for the \disklz\ sample above. We add those galaxy counts to each redshift bin of the \disklz\ sample. The completeness correction effectively increases the size of the \disklz\ sample to 525 galaxies. The bar fractions derived below are thus conservative lower limits.

\subsection{Uncertainties and Measurement Errors}\label{sec:uncertainties}

The goal of this study is to determine the fraction of the $L > L^*$ disk galaxy population at redshifts $0.5 \leq z \leq 2$ with strong bar features. We must therefore account for several potential sources of uncertainty in the measurement of the population bar fraction.

First, there are sampling errors due to the fact that we cannot sample the complete population of disk galaxies. When considering fractions of populations with a given attribute (such as a bar feature), the Normal approximation systematically underestimates proportional confidence errors when the true population fraction approaches 0 or 1, especially for small sample sizes. On the other hand, \cite{cameron11} convincingly argues that an alternative and often-used approach estimating wider confidence intervals in the case of small number statistics \citep{clopperpearson,gehrels86} systematically overestimates these confidence intervals. For that reason, \citeauthor{cameron11} advocates a Bayesian approach to binomial confidence intervals, which we adopt in this study to estimate the uncertainty due to incomplete sampling. The full 68\% confidence intervals for bar fractions in this study range from 0.04 to 0.07 around the measured fractions at each redshift. We additionally apply this method to re-calculate uncertainties for all the previous studies of bar population fractions to which we compare our results in Section \ref{sec:results}. 

There are also numerous sources of standard measurement error that could affect the measured population fractions. Here we consider the two additional sources of measurement error likely to have the greatest effect on the bar fractions: classification errors and errors in photometric redshifts.

We have estimated the uncertainties introduced into the bar fractions by photometric redshift errors via a bootstrap resampling of the redshifts in the \diskparent\ sample. Specifically, we resample the \diskparent\ sample $10^5$ times using a Monte Carlo method to vary the redshifts within the measured $\Delta z$; 92\% of the sample is assumed to vary Normally around this value, and 8\% are catastrophic outliers with errors uniformly distributed between $0.2 \leq \Delta z \leq 2$. Errors on spectroscopic redshifts are assumed to be negligible. For each resampling, we re-calculate luminosities and re-select the \disklz\ and \barlz\ samples based on the resampled luminosities and redshifts. Using this method we find the additional uncertainties introduced into the bar fraction are small: $\sigma_{frac{\rm ,~}photoz} \lesssim 0.01$.

Errors in visual classifications are greatly reduced by the Galaxy Zoo approach, which combines multiple independent classifiers. Each of the galaxies in the \diskparent\ sample have at least 40 independent classifications, enough that answers given in the first few branches of the classification tree have converged to a stable percentage of votes for a given feature \citep{willett13}. The selection method using vote percentages has also been demonstrated in previous studies to be very robust.

However, as described above, the volunteer classifications for the fourth-tier question, which directly asks about bar features (Figure \ref{fig:sampleselection}), are not yet complete and have not uniformly converged in the \diskparent\ sample. The bar classifications from 8 of the authors of this study are complete, but the smaller number may introduce additional uncertainties into the measured bar fractions. Specifically, the mean and median of the individually classified population bar fractions are fully consistent with the bar fractions described in Section \ref{sec:results}, but the spread from individual classifiers ranges from $\sigma_{frac,~class} \approx 0.04$ to $0.07$, depending on redshift bin.

Combining the errors from these three sources produces estimates of combined $1 \sigma$ uncertainties of $0.04 \leq \sigma_{frac,~comb} \leq 0.09$, which are shown in Figure \ref{fig:barfrac}. We also show the error bars from the binomial confidence intervals calculated according to \citet{cameron11} in both Figures \ref{fig:barfrac} and \ref{fig:barfrac_context}, as these error bars provide a uniform context to compare with the results from other studies. }

%
%
\section{Results: Bar fractions}\label{sec:results}
%
%

\begin{figure}
\includegraphics[scale=0.425]{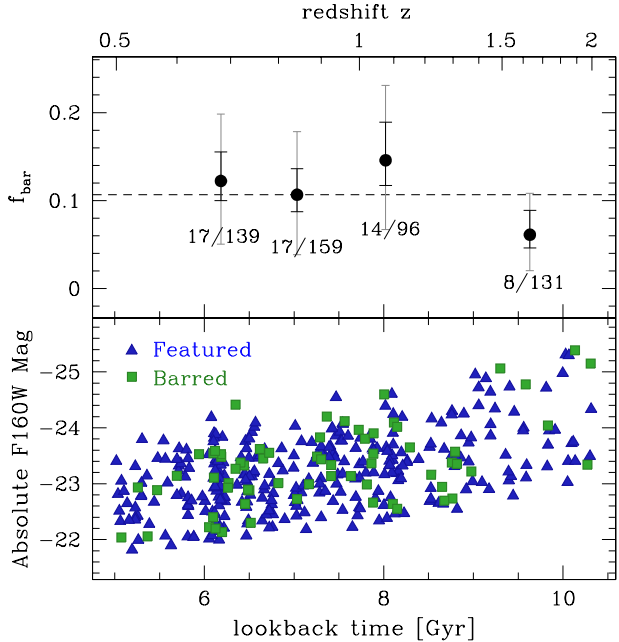}
\caption{
\emph{Top panel:} Bar fraction versus lookback time {\changed (black circles)} {\changed for the completeness-corrected \disklz\ sample}. Black error bars are 68\% {\changed Bayesian} binomial confidence intervals \citep{cameron11}; {\changed gray error bars are $1 \sigma$ uncertainties combining the binomial confidence intervals with uncertainties due to photometric redshift measurement error and classification error (described in Section \ref{sec:uncertainties}).} Within the uncertainties, the bar fraction is consistent with no evolution from $0.5 \leq z \leq 2$. Bins were chosen to enclose similar lookback time intervals; the bar fraction across all bins ({\changed $10.7^{+6.3}_{-3.5}\%$, combined errors}) is shown as a dashed {\changed line. \emph{Bottom panel:}} absolute $H$-band magnitudes of the featured disk sample from which the fractions are drawn.
}
\label{fig:barfrac}
\end{figure}

The fraction of disk galaxies with visually identified strong bars between $0.5 \leq z \leq 2$ is $\sim 10\%$, a figure that is robust to moderate changes in luminosity ranges or vote fractions for detected features, lack of clumpiness, disk inclination angle, and strong bar features. Figure \ref{fig:barfrac} shows the bar fraction with lookback time, from $t_{lb} = 5.0$~Gyr ($z = 0.5$) to 10.2~Gyr ($z=2.0$). The {\changed \disklz } sample encompasses the same subset of the galaxy luminosity function relative to the evolving $L^*${\changed . This} conservative selection to ensure detectability of features (or lack thereof) to $z = 2$ means the galaxies examined here are all brighter than $L^*$ at their epoch. 

\begin{figure}
\includegraphics[scale=0.425]{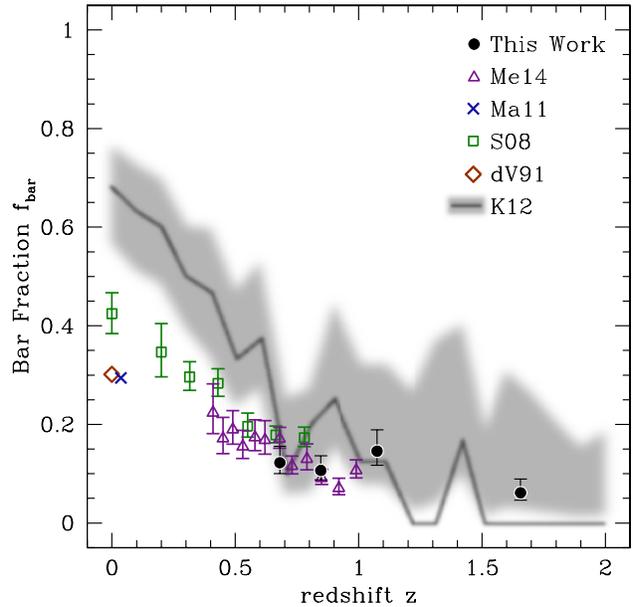}
\caption{
Fraction of disk galaxies having a strong bar feature versus redshift, in the context of other work assessing visual strong bar fraction. 
All shading and error bars indicate {\changed $1 \sigma$} Bayesian binomial confidence intervals \citep{cameron11}; {\changed where necessary, we have re-calculated uncertainties of other studies so that all uncertainties shown here are based on the same method.
Error} bars for the \citet[][Ma11, blue cross]{masters11a} and \citet[][dV91, red diamond]{RC3} fractions are smaller than the size of the points and are omitted. 
At higher redshift, bar fractions in this work (black circles) at $z < 1$ are consistent with those of \citet[][S08, green squares]{sheth08}  and \citet[][Me14, purple triangles]{melvin14} despite differences in selection methods and including our {\changed conservative completeness correction}. 
\citet[][K12]{kraljic12} computed the fraction of strong bars to $z=2$ among {\changed modelled} disk galaxies that evolved to stellar masses $M_\ast \approx 10^{10-11} \mmsun $ (shaded region); the predicted bar fraction is consistent with that observed here within the uncertainties, although we note that differences between simulated and observed mass/luminosity ranges make direct quantitative comparisons more difficult. 
}
\label{fig:barfrac_context}
\end{figure}

Figure \ref{fig:barfrac_context} shows the visually identified strong bar fraction versus redshift in the context of other work, both observational and theoretical. Within the redshift range where we overlap with other observational studies, the bar fraction is consistent. However, the bar fraction with redshift appears to flatten at $z > 1$. 

Within this sample, and given the uncertainties {\changed described in Section \ref{sec:uncertainties}}, the bar fraction is consistent with zero evolution between $1 < z < 2$. Many studies of the bar fraction at $z \lesssim 1$ find that the bar fraction does evolve, though these findings are not unanimous \citep{abraham96,abraham99,jogee04,d_elmegreen04b,d_elmegreen05,sheth08,cameron10,melvin14}. Two independent studies of the full COSMOS-ACS sample \citep{sheth08,melvin14} show that the fraction of visually identified strong bars decreases with redshift, from approximately 35\% at $z = 0.2$ to 15\% at $z = 1$. 

Using zoom-in cosmological simulations of 33 field and loose group galaxies, \citet{kraljic12} find that disk galaxies at $z \gtrsim 1$ are generally too dynamically hot to become unstable to bar formation; this manifests itself as a decreasing bar fraction with increasing redshift. Although the quantitative bar fractions in their simulations depend on the threshold used to define a bar feature, the fraction of disk galaxies hosting bars drops to zero, or near zero, by any definition they use \citep[Figure \ref{fig:barfrac_context} shows their standard ``strong bar'' definition, which is the closest to observational samples defined by visual classifications such as those here and in previous work;][]{masters11a,willett13,melvin14}. This initially appears inconsistent with our results showing a low, but non-zero, bar fraction {\changed (the observed bar fraction is formally inconsistent with 0 at the $>3 \sigma$ level)}. However,  due to the very small number of simulated galaxies in \citeauthor{kraljic12} that are disk galaxies at $z > 1$, a complete lack of bar feature detection within the subset of their sample identified as disk galaxies does not directly predict a $0\%$ bar fraction{\changed  , and given the small sample the uncertainties quoted in that study (using the Normal approximation) are likely underestimates. We have re-calculated the uncertainties quoted in \citeauthor{kraljic12}, using the Bayesian approach to compute binomial confidence intervals \citep{cameron11} discussed in Section \ref{sec:uncertainties}.} Given this approach, the lack of detection of bars at $z > 1.5$ in the simulations is consistent with a bar fraction of up to $\approx 30$\% at these redshifts, within the {\changed recalculated} 68\% confidence intervals {\changed (shown in Figure \ref{fig:barfrac_context})}. 

We also note that the galaxy masses and luminosities used in the simulations were on average lower than those examined in this work{\changed : the model galaxies in the simulations reached $M_\ast \approx 10^{10-11} \mmsun $ at $z = 0$, whereas the galaxies in the \disklz\ sample are within that stellar mass range by $0.5 \leq z \leq 2$ \citep{ilbert09, whitaker11, hartley13}. This makes a direct comparison between the simulations and} this work more difficult, as bar fraction also depends on stellar mass \citep{sheth08,melvin14}. \citeauthor{kraljic12} predict that massive disk galaxies will be more likely to form bars at higher redshift than lower mass disk galaxies due to higher-mass galaxies reaching dynamical maturity at earlier epochs. This is qualitatively consistent with our finding that the bar fraction at $z \sim 2$ may be as high as {\changed $11\%$ within $1 \sigma$ combined uncertainties (Section \ref{sec:uncertainties})}, but a direct and quantitative theoretical comparison to our observational result is currently not possible given available simulations. Expanded simulations encompassing galaxies with higher stellar masses would help to advance this field further.

Our results agree with previous work that the main epoch of disk settling (and thus bar formation) in the disk galaxy population begins at $z < 1$. However, bars are not completely absent even at $z \sim 2$: some disks at the masses probed by our sample {\changed ($10 \lesssim \log M_\ast [\mmsun ] \lesssim 11.3$)} are mature enough even by this epoch ($\sim 3-4$ Gyr after the Big Bang) to host a bar. 

Whether the bar features are analogous to long-lived bars in dynamically cold disks at lower redshift or are shorter-lived features triggered within dynamically warmer disks is unclear from examination of bar fractions alone. Examination of individual simulated galaxies by \citeauthor{kraljic12} indicates that bars formed at $z > 1.5$ tend to undergo shorter cycles of formation and destruction., and there is some evidence that short-lived grand design spiral features more commonly associated with mature disks can be triggered by interactions at $z > 2$ \citep{law12}.

Thus the incidence of bars in massive high-redshift disks may be due at least in part to galaxy interactions and mergers, combined with shorter bar lifetimes due to dynamically warmer disks. Minor galaxy mergers may dynamically heat a disk and destroy a bar, or they may trigger the formation of a bar, depending on the particulars of the interaction \citep{noguchi88,gerin90,berentzen03,berentzen04}. The relative likelihood of these contrasting end results, combined with the incidence of minor mergers among this population at $z \lesssim 2$ {\changed \citep[e.g.,][]{lotz11}}, may combine to produce a net effect that stabilizes the bar fraction at $f_{\rm bar} \sim 10$\% during this epoch of galaxy assembly. 

Among the galaxies in the highest-redshift bin of the sample, 2 of the 8 barred galaxies appear to be undergoing an interaction or merger, and another 2 appear tidally disturbed, possibly by a nearby companion. This may suggest these bar features are merger-induced; on the other hand, mergers and interactions are not particularly rare during this epoch of galaxy assembly, so their appearance in the same galaxy {\changed population} during the same epoch does not necessarily indicate a causal link. 

To investigate this further, we examined the distributions of Galaxy Zoo vote fractions for the question ``Is the galaxy currently merging or is there any sign of tidal debris?'', a second-tier question in the {\changed classification} tree to which the possible responses are ``Merger'', ``Tidal features'', ``Both'', or ``Neither''. Kolmogorov-Smirnov tests between the barred and unbarred disk galaxy samples in any redshift bin for vote fractions for responses $f_{\rm merger}$, $f_{\rm tidal}$, $f_{\rm both}$, and the sum of these fractions, are inconclusive {\changed (typical $p \sim 0.4$, with $0.08 \leq p \leq 0.92$ among the K-S tests, meaning the null hypothesis cannot be ruled out for any test)}. Resolving the question of whether shorter-lived bars are triggered by interactions and/or mergers may be possible in the future, upon the full reduction of Galaxy Zoo: CANDELS data and the addition of galaxy images from the remaining CANDELS fields to the Galaxy Zoo sample.

%
%
\section{Summary}
%
%

Using visual classifications of rest-frame optical \emph{HST} galaxy images from the ongoing Galaxy Zoo: CANDELS project, we examined for the first time the fraction of disk galaxies hosting a bar feature to $z \sim 2$ in order to trace the dynamical state of disks as early as $\sim 3$~Gyr after the Big Bang. We find that the bar fraction to $z \sim 1$ is consistent with previous studies using similar analysis methods. 

At $z > 1$, the bar fraction is approximately $10$\% and consistent with no evolution to $z \sim 2$. This is qualitatively consistent with the predictions of zoom-in cosmological simulations, although further work is needed to determine whether simulations of disk galaxies with $L > L^*$ predict the same quantitative strong bar fraction at $z < 2$. 

That the bar fraction from $0.5 < z < 2$ appears to be small but constant among massive disk galaxies implies that massive disk dynamics do not rapidly change on average over this period. Further clarification may come in the future when additional detailed morphological classifications of deep $z \sim 2$ rest-frame optical galaxy images are available{\changed . Future} comparison with independent morphologies of the same galaxies  \citep[e.g.,][]{kartaltepe14} as well as additional simulations will help provide a more nuanced understanding of the underlying physical causes of this apparently stable bar fraction. 
  
%
%
\section*{Acknowledgments}
%
%

{\changed We are grateful to the anonymous referee for helpful comments and suggestions which improved this manuscript.}
TOPCAT \citep{taylor05} and an OS X widget form of the JavaScript Cosmology Calculator {\changed \citep{wright06,simpson13}} were used while preparing this paper. 
BDS gratefully acknowledges support from the Oxford Martin School, Worcester College and Balliol College, Oxford.
TM acknowledges funding from the Science and Technology Facilities Council ST/J500665/1.
KLM acknowledges funding from The Leverhulme Trust as a 2010 Early Career Fellow.
KWW and LF acknowledge funding from the UMN Grant-In-Aid program. 
{\changed RJS acknowledges funding from the Science and Technology Facilities Council ST/K502236/1.}
RCN acknowledges STFC Rolling Grant ST/I001204/1 to ICG for ÔSurvey Cosmology and AstrophysicsÕ. 
KS gratefully acknowledges support from Swiss National Science Foundation Grant PP00P2\_138979/1. 

The development of Galaxy Zoo was supported in part by the Alfred P. Sloan Foundation. Galaxy Zoo was supported by The Leverhulme Trust. 

This work is based on observations taken by the CANDELS Multi-Cycle Treasury Program with the NASA/ESA HST, which is operated by the Association of Universities for Research in Astronomy, Inc., under NASA contract NAS5-26555.
  
\bibliographystyle{mn2e}
\bibliography{refs}

\end{document}